\documentclass[conference]{IEEEtran}
\IEEEoverridecommandlockouts
\usepackage{cite}
\usepackage{amsmath,amssymb,amsfonts}
\usepackage{graphicx}
\usepackage{textcomp}
\usepackage{style}
\usepackage{authblk}

\def\BibTeX{{\rm B\kern-.05em{\sc i\kern-.025em b}\kern-.08em
    T\kern-.1667em\lower.7ex\hbox{E}\kern-.125emX}}
\begin{document}

\title{Accelerating Maximal Biclique Enumeration on GPUs}

\author[1]{Chou-Ying Hsieh}
\author[1]{Chia-Ming Chang}
\author[2]{Po-Hsiu Cheng}
\author[1]{Sy-Yen Kuo}
\affil[1]{Department of Electrical Engineering, National Taiwan University, Taipei, Taiwan}
\affil[2]{Graduate Institute of Electronic Engineering, National Taiwan University, Taipei, Taiwan}
\affil[ ]{\textit {\{f07921043, r10921101, r10943151, sykuo\}@ntu.edu.tw}}
%\eid{123@gmail.com}

\renewcommand\Authands{ and }

%\graphicspath{{./Figures/}}

\maketitle

\begin{abstract}
Maximal Biclique Enumeration (MBE) holds critical importance in graph theory with applications extending across fields such as bioinformatics, social networks, and recommendation systems. However, its computational complexity presents barriers for efficiently scaling to large graphs. To address these challenges, we introduce cuMBE, a GPU-optimized parallel algorithm for MBE. Utilizing a unique data structure, called compact array, cuMBE eradicates the need for recursion, thereby significantly minimizing dynamic memory requirements and computational overhead. 
The algorithm utilizes a hybrid parallelism approach, in which GPU thread blocks handle coarse-grained tasks associated with part of the search process. 
Besides, we implement three fine-grained optimizations within each thread block to enhance performance. 
Further, we integrate a work-stealing mechanism to mitigate workload imbalances among thread blocks. 
Our experiments reveal that cuMBE achieves geometric mean 6.1x and 6.3x speedup compared to the state-of-the-art serial algorithm and parallel CPU-based algorithm on both common and real-world datasets, respectively.
\end{abstract}

% \begin{IEEEkeywords}
% component, formatting, style, styling, insert
% \end{IEEEkeywords}

\section{Introduction}
\label{sec:introduction}

The graph, a central discipline in computer science, servers as a versatile representation tool for complex real-world problems.
Among its diverse applications, the concept of bicliques, 
especially maximal biclique, 
stands as a core problem. 
A biclique is a complete subgraph of a biparitite graph, where every vertex in one set connects to every vertex in another set.
A maximal biclique extends this concept as a biclique that cannot include any other adjacent vertices without violating the biclique condition. 

Maximal biclique enumeration (MBE), which finds all the maximal bicliques in a given bipartite,
has profound implication across a spectrum of fields such as bioinfomatics \cite{cheng2000biclustering, tanay2002discovering, bu2003topological, li2006discovering, voggenreiter2012exact}, text mining \cite{muhammad2016summarizing, shaham2016finding, yoshinaka2011towards}, recommendation systems \cite{maier2021biclique, maier2022bipartite}, and even the accelerating of the graph neural network \cite{jia2020redundancy}.
In these field, the maximal biclque can reveal intricate structure and essential relationships.
For instance, in recommendation systems, 
a maximal biclique can represent a group of users who share high preference on certain products.
This property can help recommendation systems be more accurate. 

One of the most prestigious MBE algorithm is the MBEA proposed by Zhang, et.al \cite{zhang2008finding}.
Inspired by Bron-Kerbosch's maximal clique enumeration (MCE) \cite{bron1973algorithm},
MBEA refines the backtracking technique in Bron-Kerbosch's algorithm by traversing a searching tree. 
During the traversal, the procedure adds a vertex into the current biclique and check if it can construct a maximal biclique recursively. 
Unfortunately, due to the inherent combinatorial complexity of MBE, the computational task is often intensive and even becomes significantly challenging for large-scale real-world graph data. 
Despite there are numerous prior works \cite{abidi2020pivot, ooMBE, das2019shared, zhang2014finding, liu2006efficient} have introduced versatile improvements on reducing the algorithm complexity, 
the state-of-the-art \cite{chen2022efficient} sequential approach still consumes considerable time on large-scale datasets.
To address the limitation, Das et al. \cite{das2019shared} leveraged the power of multi-core CPUs to parallelize the MBE algorithm, achieving significant improvements in performance. 
Despite these advancements, the pursuit for greater scalability persists. 
The emergence of Graphics Processing Units (GPUs), renowned for their extensive parallel computational capabilities, presents an intriguing prospect for further enhancing the scalability beyond what has been achieved with CPUs alone
Nevertheless, the specific application of GPUs for efficient maximal biclique enumeration remains an open problem.
In particular, avoiding memory explosion on GPUs, managing the granularity of parallelism, 
and distributing workload evenly across GPU threads pose unique challenges. 

In this paper, we present \emph{cuMBE}\footnote{GitHub repository: https://github.com/NTUDDSNLab/cuMBE}, a novel GPU parallel algorithm for MBE. 
cuMBE employs a depth-first approach and addresses its recursive overhead with an innovative data structure, the \emph{compact array}. This mitigates memory overload and negates the need for dynamic memory allocation. 
To efficiently manage tasks, cuMBE adopts \emph{hybrid parallelism}, with a GPU thread block handling coarse-grained tasks and optimized fine-grained task management within each block. 
To further enhance performance, a \emph{k-level work-stealing algorithm} is incorporated to balance workload among thread blocks. 
Comparative analysis against traditional CPU-based methods, using real-world graph datasets, 
demonstrates cuMBE's superior performance with an average 6.1x and 6.3x speedups over state-of-the-art serial and parallel algorithms, respectively, on prevalent datasets. 
We also demonstrate that the implemented work-stealing algorithm effectively mitigates workload disparities.
Notably, the cuMBE code is openly available to facilitate future research.

The paper is organized as the below:
In the next section, we first introduce the preliminary of MBE algorithm as well as the advanced techniques used in MBE algorithm. 
Then, we point out the challenges that implements MBEA on GPUs and introduce cuMBE features in Section \ref{sec:methodologies}. 
In Section \ref{sec:evaluation}, we evaluate and analyze cuMBE. 
Lastly, we provide some prior works related to cuMBE and summarize the paper.

\section{Background}
\label{sec:background}

\subsection{Maximal Biclique Enumeration}

\begin{figure}[t]
    \centering
    \includegraphics[width=0.9\linewidth]{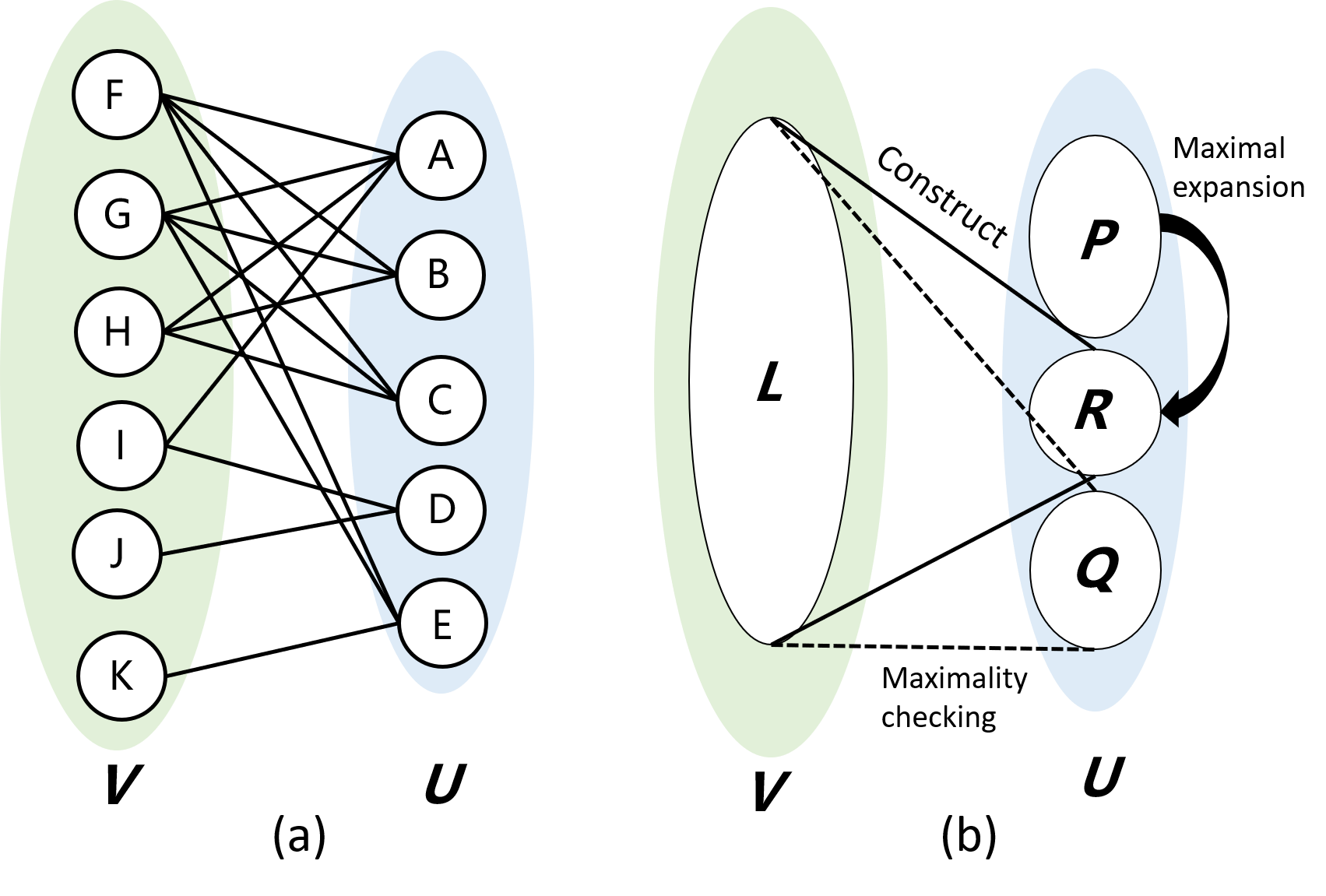}
    \caption{(a) Example bipartite graph. (b) The relationship of four sets in the state-of-the-art MBEA. The $P$ set stores the candidate vertex to be added to $R$, while the $R$ set induces the $L$; the $Q$ set check the maximality of $L$, while the $R$ set expands itself to maximal by moving vertices from $P$ to $R$.}
    \label{fig:bipartite}
\end{figure}

Let $G = (U \cup V, E)$ be a \textbf{bipartite graph}, where $U$ and $V$ are two partitioned sets of vertices.
$E \subseteq U \times V $, which is the edge set of $G$. 
Without loss of generality, we assume that the $|V| \geq |U|$.
Figure \ref{fig:bipartite} (a) shows a bipartite graph with $U=\{A, B, C, D, E\}$ and $V=\{F, G, H, I, J, K\}$.
A \textbf{biclique} shown in Figure \ref{fig:bipartite} (b) $H = (L \cup R, E')$ in the bipartite graph $G$, also known as the complete bipartite subgraph, 
satisfies that $L \subseteq V$, $R \subseteq U$, 
and $\forall u \in L, \forall v \in R, (u, v) \in E'$.
A biclique is \textbf{maximal} if it is not a proper subset of any other biclique. 
In other words, there are no additional vertices outside the biclique that can be added to either $L$ or $R$ without violating the definition of a biclique.
For instance, $B_{1} = (\{ A,B,C \} \cup \{ F,G,H \})$ and $ B_{2} = (\{ A,B\} \cup \{ F,G,H \})$ are two bicliques in Figure \ref{fig:bipartite} (a),
but only $B_{1}$ is maximal since we can expand $B_{2}$ with the vertex $C$ without removing any vertices in $\{ F, G, H\}$.
The maximal biclique enumeration (MBE) is the algorithm to report all the maximal bicliques in a given bipartite graph.

To explain the MBE algorithm clearly, we first define some terminologies. 
The set of \emph{neighbors} for a vertex $v \in L$ or $v \in R$ is denoted as $N(v)$; given a vertex set $X \subseteq L$ or $X \subseteq R$, we use $N(X)$ to denote the \emph{common neighbors} of vertices in $X$.
The mathematical definition of common neighbor is:
$N(v) = \{u | (v, u) \in E\}$ and $N(X) = \bigcap_{\forall x \in X} N(x) $.
We also denote $N_{2}(v)$ and $N_{2}(X)$ as the \emph{2-hop neighbors} of the vertex $V$ and the vertex set $X$, respectively.
The 2-hop neighbor of vertex $v$ and vertex set $X$ are defined as: $N_{2}(v) = N(N(v))$ and $N_{2}(X) = N(N(X))$.
We also denote $N(v, H)$ as $N(v) \cap H$, where $H \subseteq G$.

\subsection{The State-of-the-art MBEA}
\label{sec:mbea}

\begin{algorithm}[t]
\caption{MBEA($L, R, P, Q$)}
\label{algo:MBEA}
\DontPrintSemicolon
  
\KwInput{$L$: the subset of $V$, $R$: the subset of $U$, $P$: the candidate vertex set, $Q$: the set of maximality checking}
\KwOutput{$B$: set of all maximal bicliques}
\KwData{$L' \leftarrow \phi$, $R' \leftarrow R$, $P' \leftarrow \phi$, $Q' \leftarrow \phi$, $x$: the candidate vertex}

\tcc{Step 0: Embedded pruning}
\While{$|P| > 0$} {
    \tcc{Step 1: Candidate selection}
    $x = P\text{.pop()}$ \;
    $R' = R \cup \{ x \}$ \;
    
    \tcc{Step 2: $L'$ construction}
    \ForEach{$v \in L$ and $(x, v) \in E$} {
        $L' = L' \cup \{ v \}$
    }

    \tcc{Step 3: Maximality checking}
    $\text{isMaximal} \leftarrow \text{True}$ \;
    
    \ForEach{$v \in Q$} {
        \If{$|N(v) \cap L'| = |L'|$} {
            $\text{isMaximal} \leftarrow \text{False}$ \;
            \Break \;
        }
        \ElseIf{$|N(v) \cap L'| > 0$} {
            $Q' = Q' \cup \{ v \}$
        }
    }
    
    \If{$\text{isMaximal} = \text{True}$} {
        \tcc{Step 4: Maximal expansion}
        \ForEach{$v \in P$}{
            \If{$|N(v) \cap L'| = |L'|$} {
                $R' = R' \cup \{ v \}$
            }
            \ElseIf{$|N(v) \cap L'| > 0$} {
                $P' = P' \cup \{ v \}$
            }
        }
        $B = B \cup \{ (L', R') \}$ \tcp*{Add a maximal biclique}
        \If{$|P'| \neq 0$} {
            MBEA($L'$, $R'$, $P'$, $Q'$) \tcp*{Recursive call}
        }
    }
    $Q = Q \cup \{ x \}$ \tcp*{Move the tested vertex to $Q$}
}

\end{algorithm}

\begin{figure*}
    \centering
    \includegraphics[width=0.8\textwidth, height=15cm]{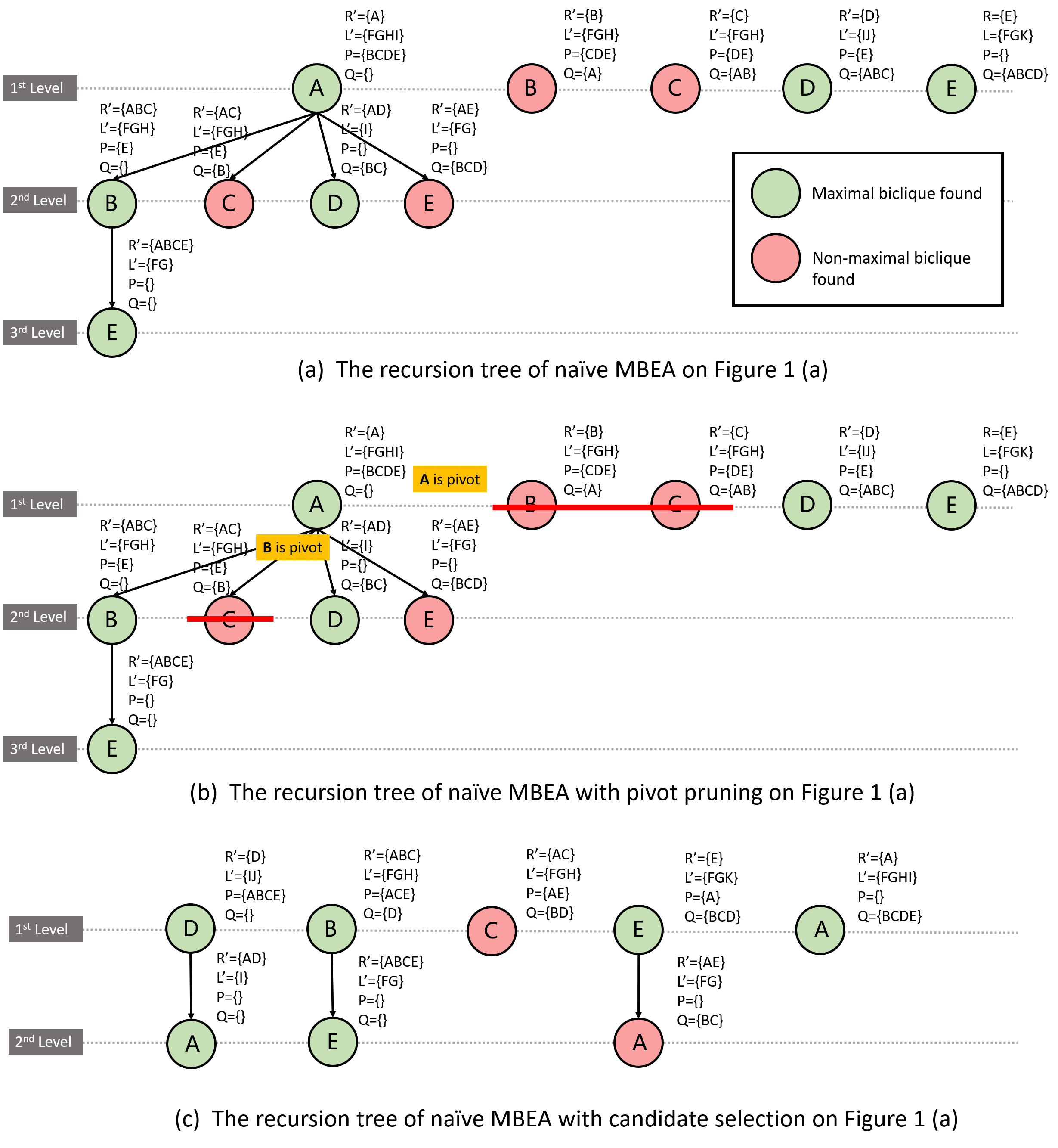}
    \caption{The MBEA with different advanced techniques on the given example bipartite shown in Figure \ref{fig:bipartite}. Each circle represents a checking process. The letter inside a circle is the vertex $x$ selected in that process. The maximal biclique found is represented as $B = \{R' \cup L'\}$.}
    \label{fig:recursion_tree}
\end{figure*}

Zhang, et al. proposed the MBEA \cite{zhang2008finding}, which is the first serial algorithm designed specifically for MBE. 
It uses the branch-and-bound approach to explore the search space of possible bicliques in a bipartite recursively, shown in Algorithm \ref{algo:MBEA}.
This algorithm focuses on the four sets, $L$, $R$, $P$, and $Q$. 
The relation of these sets is shown in Figure \ref{fig:bipartite}.(b).
Fundamentally, MBEA systematically investigates all potential bicliques, where $R, P, Q \subseteq U$ and $L \subseteq V$. 
At first, the procedure checks whether the vertex $x$ from the \emph{candidate set} $P$ forms a maximal biclique or not. 
If a maximal biclique is found, 
this procedure will recursively search the next biclique by interacting with the given four sets $L$, $R$, $P$, $Q$.
The recursive progress can represent as a recursive tree shown in Figure \ref{fig:recursion_tree}.
The MBEA algorithm begins with iterating all the candidate vertices in $P$ (line 1).
If $P$ is not empty, the procedure will pop out a vertex $x$ and 
insert it to $R'$, which is potentially part of maximal biclique in $U$.
After the insertion, MBEA creates the $L'$ by testing if the vertex in the $L$ set also connects to $x$ (lines 4-5).
With both $R'$ and $L'$, MBEA then checks the maximality of $L'$ with the $Q$ (line 7-12). 
Since the $Q$ set collects tested vertices that cannot comprise a maximal biclique, 
the common neighbor of any vertices in $Q$ cannot fully connect to $L'$.
Otherwise, these vertices should be added to form a maximal biclique.
The procedure uses the size of the two set $N(v) \cap L'$ and $L'$ to check if the two sets are equal. 
The rest of the vertex in $Q$ which has at least one common neighbor with $L'$ is added to $Q'$ for the next level searching.
Once we ensure the maximality of $L'$, 
we can use $L'$ to induce the maximal biclique by expanding $R'$ to maximal (lines 14-17).
Last but not least, if there is some vertex in the next-level candidate set $P'$, 
the procedure will further search $L', R', P'$, and $Q'$ recursively.

\subsection{Advanced Techniques on MBEA}
\label{sec:techniques}

Prior works \cite{abidi2020pivot} have introduced different techniques to improve MBEA efficiency. 
We arrange and conclude these techniques in three main aspects: 
embedded pruning, candidate selection, and parallel searching. 
The first two aspects focus on reducing the complexity of the algorithm itself, while the last aspect accelerates the computation of MBEA by separating independent tasks on multiple computing units.

\textbf{Embedded pruning techniques.} The runtime complexity of iMBEA is closely tied to the depth of recursion, primarily determined by the size of set $P$.
Initially, brute-force approach would test all vertices in $P$ individually,
generating a plethora of inefficient branches.
Previous work has introduced pruning techniques to mitigate this, employing a concept called the Neighbor Containment Relationship (NCR) \cite{abidi2020pivot, das2019shared, ooMBE, liu2006efficient}.
Formally, we define $NCR(v, v', H)$ as the vertex $v$ containing $v'$ if $N(v', H) \subseteq N(v, H)$ for a subgraph $H \subseteq G$.
This property allows for the removal of $v$ from $P$, 
thereby reducing the search space while preserving the quality of results. 
Notably, Zhang et al. first employed this concept during the maximal biclique expansion process \cite{zhang2014finding}, 
removing $v$ from $P$ if it did not connect to any vertex in the set $L$ - $L'$.
Building on Bronn and Kerbosh's \emph{pivot pruning} \cite{bron1973algorithm}, 
Abidi et al. proposed PMBEA \cite{abidi2020pivot}, which uses a Containment Directed Acyclic Graph (CDAG) to refine $P$ by partitioning it into two sets, 
$C$ and $C'$, based on a pivot vertex $v_{p}$.
Chen et al. further optimized this approach by introducing batch pivot selection \cite{chen2022efficient}. Figure \ref{fig:recursion_tree}(b) illustrates how pivotal pruning effectively trims the search space.
Notably, those pruning techniques are orthogonal to our work.
However, the effectiveness of these techniques relies heavily on the choice of pivots made by different heuristic algorithms. Picking the right vertices, like vertex $A$ and $B$, is crucial for performance gains; 
otherwise, finding suitable pivots could be time-consuming. 
As a result, we chose not to include these pruning techniques in our implementation, focusing instead on the core optimization of cuMBE on GPUs.

\textbf{Candidate selection techniques.} 
% TODO: Point out the difference between pruning and ordering
Unlike embedded pruning, which directly eliminates candidate vertices from the set $P$ to reduce the search space, the order in which vertices are selected can also influence the efficiency of the branch-and-bound search. 
This concept is illustrated in Figure \ref{fig:recursion_tree}. 
iMBE \cite{zhang2014finding}, the first paper to introduce this idea, opts to select vertices from $P$ based on the ascending size of their common neighbor set, leading to notable efficiency gains.
One reason this ordering enhances performance is related to \emph{maximal expansion}, as depicted in Figure \ref{fig:recursion_tree}.
During this phase, the current biclique is expanded to its maximum size by adding vertices from $P$. 
This means a vertex with more common neighbors will inherently include those with fewer, 
leading to redundant checks and thus inefficiency.
Recent work has further refined this ordering strategy, 
offering even better time complexity for sparse graphs. 
Because this ordering method not only improves efficiency but also balances the workload among different recursion subtrees, 
we have adapted and optimized it for GPUs, which we will discuss in detail in a later section \ref{sec:fine-grained}.

\subsection{Architecture and Execution Model on GPUs}

Graph processing unit (GPU) is a massively parallel processor that utilizes many processing units operating concurrently to perform computations. 
Specifically, a modern GPU comprises several \emph{streaming processors (SMs)} working in a highly parallel, multi-threaded environment.
Multiple threads run the same program on different data points concurrently.
A SM organizes its threads into groups called \emph{warps} which execute instructions in a SIMD (Single Instruction, Multiple Data) fashion, 
namely threads inside a warp execute the same instruction with different data at the same time.
Multiple warps comprise a \emph{thread block (TB)} where threads inside a TB share a fast cached memory called shared memory.
The TB can dramatically hide the data transmission time by context switching among warps.
Based on these two features,
we utilize TBs to achieve coarse-grained parallelism, where larger tasks are performed concurrently. 
Meanwhile, within each TB, the warps handle fine-grained parallelism, where these larger tasks are further divided into SIMD task for parallel computing.

\section {Parallel MBEA on GPUs}
\label{sec:methodologies}

\subsection{Challenges and Implementation Overview}
\label{sec:challenges}

Initially, we considered a Breadth-First Search (BFS) approach to parallelize the search for maximal bicliques on GPUs. 
BFS, widely used in various graph search algorithms \cite{harish2007accelerating, busato2015efficient, vineet2009fast, hsieh2023decentralized, ji2022ispan, soman2010fast}, is naturally parallelizable but presents significant challenges for MBEA. 
Chief among them is excessive memory consumption,
a major concern given the limited on-chip memory of GPUs. 
While methods like GPU oversubscription  \cite{li2019framework, ganguly2021adaptive, ganguly2020adaptive} can partly alleviate this issue, they require slow CPU-to-GPU data transfers.

To circumvent these challenges, we employ a Depth-First Search (DFS) approach, which is considerably more memory-efficient and also widely used in other graph algorithms on GPUs \cite{chen2022efficient, almasri2022parallel, almasri2022parallelizing, chen2021sandslash}. Despite its benefits, 
DFS introduces its own set of challenges for GPU implementation. 

First, the conventional approach uses \emph{dynamic parallelism} \cite{merrill2011high, jones2012introduction, wang2014characterization} to enable recursion on GPUs.
It provides the ability to launch new kernels from existing ones, but this approach introduces substantial kernel launching overhead \cite{chen2017effisha, chen2015free, wang2010kernel}.
Second, we have to use \emph{dynamic memory allocation} to generate the next-level sets.
Although there are numerous works dedicating to accelerate the dynamic memory allocation \cite{gelado2019throughput, springer2019dynasoar, winter2020ouroboros}, 
it is still expensive compared to the static memory allocation \cite{winter2021dynamic}.
Moreover, in scenarios where a recursion tree's height was considerable, 
the memory required could still exceed what the GPU could accommodate.
Third, we face the \emph{workload imbalance} across subtrees.
This imbalance comes from the considerable disparity in the height of each subtree and the overly coarse-grained task granularity associated with the first level subtree.

To tackle these hurdles, we introduce an efficient, workload-balanced MBEA that incorporates three main optimizations: 
a compact array structure, a dynamic k-level work-stealing algorithm, and a hybrid parallelism model. 
Our compact array optimization eliminates recursion and significantly reduces dynamic memory usage. 
The work-stealing algorithm addresses workload imbalance and refines the efficiency of the DFS approach. 
Lastly, we introduce optimizations at both coarse-grained and fine-grained levels to fully exploit the GPU's computing power. 
To the best of our knowledge, cuMBE is the first work to design and implement MBEA on GPUs.

\subsection{Recursion Elimination with Compact Array}
\label{sec:data_structure}

\begin{figure}
    \centering
    \includegraphics[width=0.95\linewidth, height=5cm]{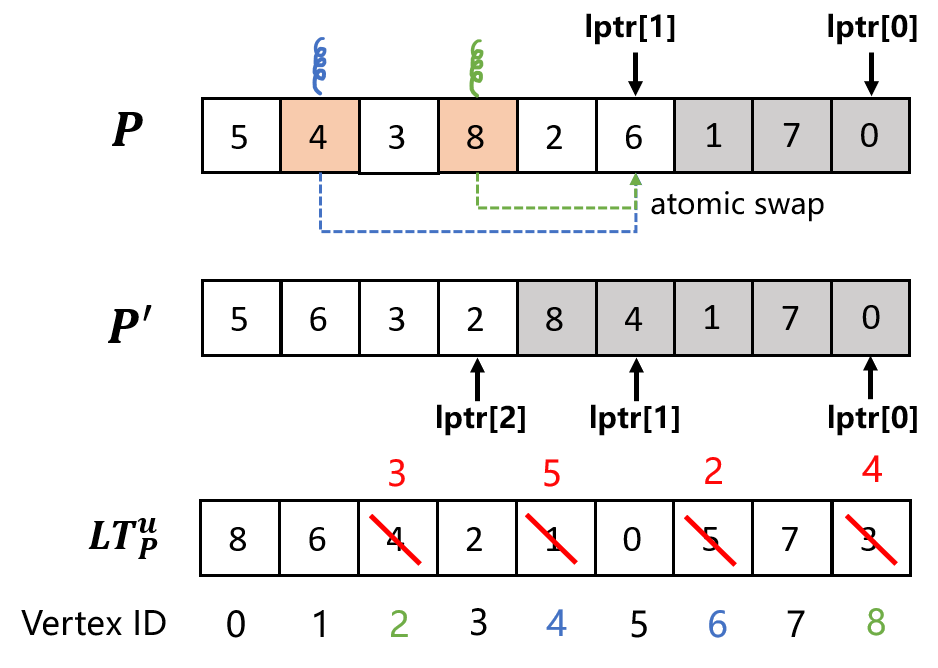}
    \caption{The example of moving vertices using compact array and lookup table in the candidate set $P$. The $P'$ is the next-level candidate set. Assuming that the blue thread executes the atomic swap first, it is responsible for updating vertex $4$ and $6$'s position in the lookup table with the index colored in blue. }
    \label{fig:compact_stack}
\end{figure}

Traditionally, eliminating the overhead of repeated kernel launches in dynamic parallelism involves converting a recursive program into an iterative one, often by using a stack to store local variables. However, dynamic memory allocation for the stack can be time-consuming and prone to memory overflow. 
To overcome these issues, we introduce a novel data structure called a \emph{compact array} to replace both recursion and dynamic memory allocation.

Our observations show that all vertices are sourced from the same graph; 
namely, the local sets, $P, Q, R, L$ in each recursive level are the subset of either $U$ or $V$.
Further, $P$ and $L$ decrease in size while $R$ increases as we delve into deeper levels of recursion. 
Although $Q$ varies, we discuss its handling in a later section. Utilizing these observations, we use pointers at each level to designate which vertices belong to current local sets, 
thereby representing each set with a single copy. 
This process is illustrated in Figure \ref{fig:compact_stack}.

According to Algorithm \ref{algo:MBEA}, $P$ is the subset of $U$;
hence, We first allocate an array with size $|U|$, where each element is the vertex in $U$.
Then, we set a level pointer called $lptr[x]$ at each level $x$ for indicating the last element of the current local set;
namely, the vertex whose index is smaller than the current level pointer belongs to the current local set.
Take Figure \ref{fig:compact_stack} for example,
the $P$ set is in the first recursive level with $lptr[1]$ and $lptr[0]$.
The vertices with white color are used in the first, while the grey vertices belong to the previous level.
Before we descend to explore maximal bicliques in the second level, 
we have to well prepare $P'$.
The $P'$ shrinks from $P$ during the candidate selection and maximal expansion process.
Assuming two threads want to move vertex $4$ and $8$ to $R'$ during the maximal expansion; 
hence, we should swap these vertices to the end of $P$ and decrement $lptr[1]$ from 5 to 3.

To parallelly swap multiple vertices to the end without race condition, 
we use an auxiliary array and an atomic counter, whose value is initialized as the current level pointer.
Threads atomically acquire and decrement the counter, while they put the vertex to the auxiliary array according to the index they acquire.
After all threads finish moving to the auxiliary array, 
we parallelly swap the vertex in this array to $P'$ and set the new level pointer $lptr[2]$ with the value of the counter.  
We also apply the compact array approach to $R, L$, and $Q$ sets to avoid memory exploration.
We can reduce the space complexity from $O(h*N)$ to $O(N)$, where $h$ is the height of the recursion tree and $N$ is the total vertex number.

\subsection{Coarse-grained Parallelism on Independent Subtrees}
\label{sec:coarse-grained}

In Section \ref{sec:mbea}, we have shown that the branch-and-bound approach of MBEA can transform to the searching of a whole recursion tree.
We observe that subtrees of this recursion tree are independent;
namely, we can execute each subtree parallelly.
We treat the $1^{st}$ level subtree as the coarse-grained task and assign a TB to handle it. 
Since a TB has its own cache memory and multiple warps for hiding the memory transfer latency with concurrent execution, 
prior works \cite{almasri2022parallelizing} also adapt the same concept on parallelizing their DFS applications.
To fulfill this coarse-grained parallelism, 
we maintain a global candidate set $P_{g}$ with compact array and a $size$ variable.
The $P_{g}$ is initialized as the $U$ set. 
Once a TB fetches a coarse-grained task, 
it first atomically decrease the $size$ by 1.
If the return value $r$ of the atomic function is larger than 1, 
which means the $P_{g}$ is non-empty, 
the TB will copy $P_{g}[0:r]$ to its local candidate set $P_{p}$ stored in the shared memory, 
which is called subtree fetching. 
This step is necessary since the shared memory has up to 100x speedup on memory access compared to the global memory.
Once the transfer is done, 
this TB can operate on  its own $P_{p}$ set independently.

\subsection{Work-stealing on K-level Independent Subtrees}
\label{sec:work-stealing}

\begin{algorithm}[t]
\DontPrintSemicolon
  
\KwInput{$k$: the maximum depth to execute work-stealing, $tid$: the thread block ID, $cur\_level$: the recursive level the thread block is at}
\KwData{$glevel \leftarrow 1$, $isPause \leftarrow False$, $ntid \leftarrow (tid + 1) \% \ ntb$}

\tcc{Determine if the thread block should enter work-stealing}
\If{$cur\_level \neq glevel$ \textbf{or} $glevel \geq k$} {\Return}

\tcc{Victim thread blocks}
\If{$isPause$} { 
    Update($P_{g}^{tid}$) \;
}

\tcc{Theif thread block}
\ElseIf{$isEmpty(P_{g})$} {
    $isPause \leftarrow True$ \;
    $glevel \leftarrow glevel + 1$ \;
}
grid.sync() \;
AtomicFetch($P_{g}$)

\caption{Work-stealing(k, tid, cur\_level)}
\label{algo:work-stealing}
\end{algorithm}

Recall from Section \ref{sec:challenges} that there is severe workload imbalance during the MBE depth-first procedure on GPUs. 
In Section \ref{sec:coarse-grained}, we assign a TB an independent $1^{st}$ level subtree for the depth-first enumeration.
However, when a TB finish its subtree searching but there is no $1^{st}$ level subtree left,
this TB have to wait until other TBs finish.
To make matter worse, the straggler TB takes up to 21.4x more time than others in some certain cases of our experiments.
To alleviate this imbalance, we design a k-level work-stealing algorithm which enables the idle TB to steal subtrees from others shown in Algorithm \ref{algo:work-stealing}.
The $k$ variable represents the maximum level of work-stealing we allow;
namely, the larger $k$ means that the TB can steal more fine-grained task.

Fundamentally, we call the work-stealing procedure every time at the end of the \texttt{While} loop in Algorithm \ref{algo:MBEA}.
Before a TB pop a vertex from the candidate set $P$, 
we first evaluate if the TB should enter the work-stealing procedure with \texttt{cur\_level} and \texttt{glevel}.
The former indicates the next level of subtree that the caller TB will fetch.
The \texttt{cur\_level} increments after the TB recursively executes the next-level subtree, 
while it decrements after the TB returns from the recursive call.
The \texttt{glevel} is a global variable which shows the level that the global work-stealing procedure works at.
We initialize it to 1 so that each TB will start to fetch $1^{st}$ level subtrees.
The \texttt{glevel} increments to 2 when there is no $1^{st}$ level subtrees.
We start the work-stealing procedure by determining if the \texttt{cur\_level} equals to the \texttt{glevel};
namely, the TB want to execute the \texttt{glevel}'s subtree.
If there is no such subtree, 
this TB then raises the \texttt{isPause} and increments the \texttt{glevel} (lines:5 - 7).
The TB which is called \textbf{theif TB} then idles until other \textbf{victim TBs} finish their ongoing $glevel^{th}$ level subtree.
These victim TBs then execute to line: 3 and push their tasks to the their own global set $P_{g}^{tid}$, where \texttt{tid} denotes as the thread block ID. 
We extend the single $P_{g}$ mentioned in Section \ref{sec:coarse-grained} to multiple sets $\{P_{g}^{tid} \, | \, tid \in [0, ntb)\}$, 
where \texttt{ntb} denotes as the number of thread block.
\texttt{Update($P_{g}^{tid}$)} makes the victim TB to sort its private candidate set $P$ set and copy the $P$ to the $P_{g}^{tid}$.
After this transfer, we can know the status of each TB. 
Once all victim TBs have updated, 
a TBs can fetch a task from $P_{g}$ by copy a candidate set from $P_{g}$ to its private $P_{p}$ (Algorithm \ref{algo:work-stealing}: line: 9).

To avoid the race condition on fetching tasks,
the TB whose ID is $tid$ should atomically decreases the $lptr[glevel]$ of the $P_{g}^{tid}$ first.
If the return value $R$ of the atomic function is larger than 0, 
the TB can safely copy $P_{g}^{x}[0:R]$ to its local candidate set.
We use the circular stealing method similar to \cite{chase2005dynamic} to steal other TB's task;
namely, if $P_{g}^{tid}$ is empty, the TB will try to steal from the $P_{g}^{tid+1}$.

Theoretically, the $k$ value can be any positive integers.
Although larger $k$ can make the workload more balanced, 
the cost of synchronization among thread blocks might increases.
In addition, the improvement of work-stealing degrades through the task becoming more fine-grained.
In Section \ref{sec:evaluation}, we discuss how to select an appropriate $k$ validated by sufficient experiments.

\begin{figure}
    \centering
    \includegraphics[width=0.95\linewidth]{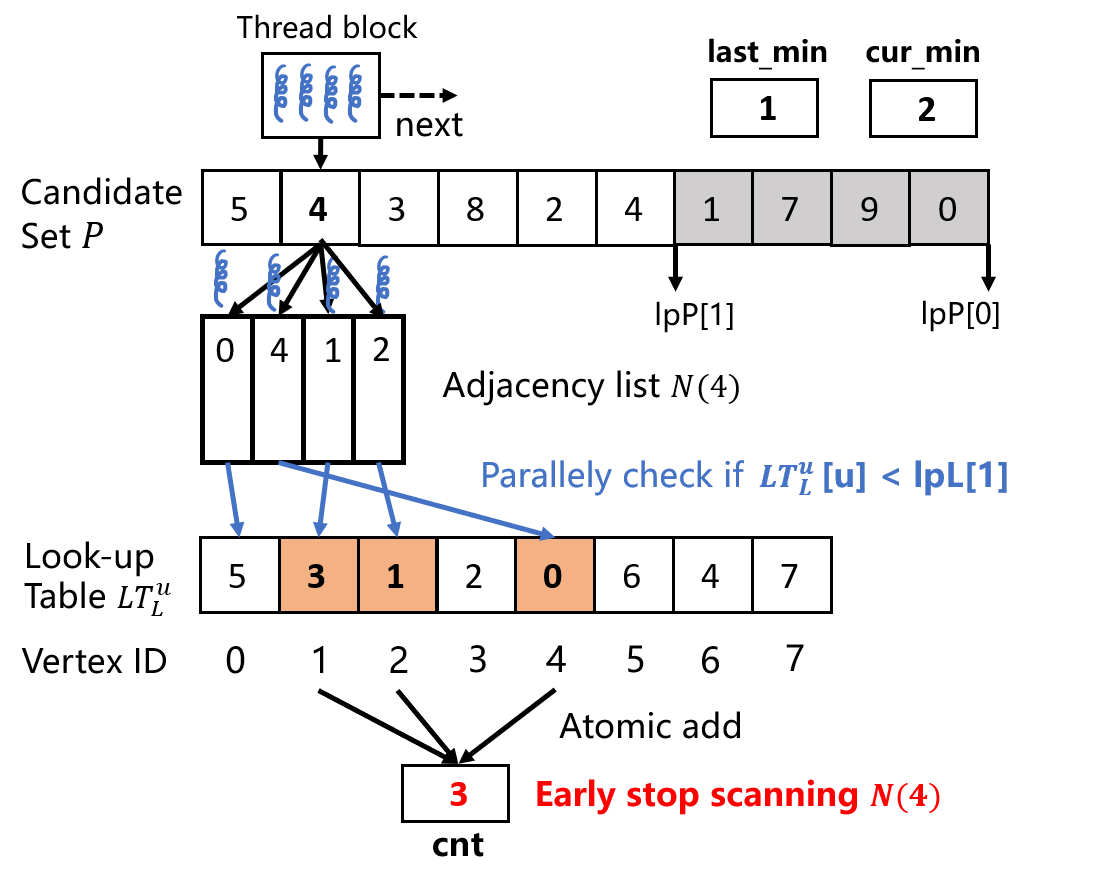}
    \caption{The early-stop technique in the candidate selection phase with the lookup table $LT^{u}_{L}$ for fast determining if the neighbors vertex $u$ is in the $L$ set.}
    \label{fig:candidate_selection}
\end{figure}

\begin{table*}[t]
\caption{Performance evaluation of maximal biclique enumeration algorithms across various datasets}
\label{table:performance}
\resizebox{0.95\linewidth}{!}{%
\begin{tabular}{|l|r|r|r|r|r|r|rr|r|rrr|}
\hline
\multicolumn{1}{|c|}{\multirow{2}{*}{Graph}} & \multicolumn{1}{c|}{\multirow{2}{*}{|U|}} & \multicolumn{1}{c|}{\multirow{2}{*}{|V|}} & \multicolumn{1}{c|}{\multirow{2}{*}{|E|}} & \multicolumn{1}{c|}{\multirow{2}{*}{\begin{tabular}[c]{@{}c@{}}Edge\\ Density\end{tabular}}} & \multicolumn{1}{c|}{\multirow{2}{*}{\begin{tabular}[c]{@{}c@{}}\# of maximal\\  biclique\end{tabular}}} & \multicolumn{1}{c|}{\multirow{2}{*}{\begin{tabular}[c]{@{}c@{}}ooMBE \\ execution\\  time (s)\end{tabular}}} & \multicolumn{2}{c|}{ParMBE execution time (s)}                                                                                                                                                     & \multicolumn{1}{c|}{\begin{tabular}[c]{@{}c@{}}cuMBE \\ execution \\ time (s)\end{tabular}} & \multicolumn{3}{c|}{cuMBE Speedup}                                                                                                                                                                                                              \\ \cline{8-13} 
\multicolumn{1}{|c|}{}                       & \multicolumn{1}{c|}{}                     & \multicolumn{1}{c|}{}                     & \multicolumn{1}{c|}{}                     & \multicolumn{1}{c|}{}                                                                        & \multicolumn{1}{c|}{}                                                                                   & \multicolumn{1}{c|}{}                                                                                        & \multicolumn{1}{c|}{\begin{tabular}[c]{@{}c@{}}Intel Xeon 8375C  \\ with 128 threads\end{tabular}} & \multicolumn{1}{c|}{\begin{tabular}[c]{@{}c@{}}AMD EPYC 7R13\\ with 192 threads\end{tabular}} & \multicolumn{1}{c|}{RTX 3090}                                                               & \multicolumn{1}{c|}{\begin{tabular}[c]{@{}c@{}}over\\ ooMBE\end{tabular}} & \multicolumn{1}{c|}{\begin{tabular}[c]{@{}c@{}}over 128\\  threads\end{tabular}} & \multicolumn{1}{c|}{\begin{tabular}[c]{@{}c@{}}over 192\\  threads\end{tabular}} \\ \hline
DBLP-author                                  & 5,624,219                                 & 1,953,085                                 & 12,282,059                                & 2.2E-06                                                                                      & 4,899,032                                                                                               & 18.78                                                                                                        & \multicolumn{1}{r|}{6.03}                                                                          & 5.46                                                                                          & 228.30                                                                                      & \multicolumn{1}{r|}{0.08}                                                 & \multicolumn{1}{r|}{0.03}                                                        & 0.02                                                                             \\ \hline
DBpedia\_locations                           & 53,407                                    & 172,079                                   & 293,697                                   & 6.4E-05                                                                                      & 75,360                                                                                                  & 0.26                                                                                                         & \multicolumn{1}{r|}{0.21}                                                                          & 0.16                                                                                          & 0.11                                                                                        & \multicolumn{1}{r|}{2.36}                                                 & \multicolumn{1}{r|}{1.91}                                                        & 1.46                                                                             \\ \hline
Marvel                                       & 12,942                                    & 6,486                                     & 96,662                                    & 2.3E-03                                                                                      & 206,135                                                                                                 & 2.67                                                                                                         & \multicolumn{1}{r|}{0.79}                                                                          & 0.58                                                                                          & 0.07                                                                                        & \multicolumn{1}{r|}{38.14}                                                & \multicolumn{1}{r|}{11.29}                                                       & 8.22                                                                             \\ \hline
YouTube                                      & 94,238                                    & 30,087                                    & 293,360                                   & 2.1E-04                                                                                      & 1,826,587                                                                                               & 40.03                                                                                                        & \multicolumn{1}{r|}{21.80}                                                                         & 13.03                                                                                         & 2.22                                                                                        & \multicolumn{1}{r|}{18.03}                                                & \multicolumn{1}{r|}{9.82}                                                        & 5.87                                                                             \\ \hline
IMDB-actor                                   & 896,302                                   & 303,617                                   & 3,782,463                                 & 2.8E-05                                                                                      & 5,160,061                                                                                               & 82.17                                                                                                        & \multicolumn{1}{r|}{157.32}                                                                        & 92.16                                                                                         & 31.47                                                                                       & \multicolumn{1}{r|}{2.61}                                                 & \multicolumn{1}{r|}{5}                                                           & 2.93                                                                             \\ \hline
stackoverflow                                & 545,195                                   & 96,678                                    & 1,301,942                                 & 4.9E-05                                                                                      & 3,320,824                                                                                               & 360.15                                                                                                       & \multicolumn{1}{r|}{1,529.22}                                                                      & 1,424.45                                                                                      & 350.07                                                                                      & \multicolumn{1}{r|}{1.03}                                                 & \multicolumn{1}{r|}{4.37}                                                        & 4.07                                                                             \\ \hline
BookCrossing                                & 340,523                                   & 105,278                                   & 1,149,739                                 & 6.4E-05                                                                                      & 54,458,953                                                                                              & 3,243.87                                                                                                     & \multicolumn{1}{r|}{2568.77}                                                                       & 1,602.07                                                                                      & 622.08                                                                                      & \multicolumn{1}{r|}{5.21}                                                 & \multicolumn{1}{r|}{4.13}                                                        & 2.58                                                                             \\ \hline
corporate-leadership                         & 24                                        & 20                                        & 99                                        & 4.1E-01                                                                                      & 66                                                                                                      & 0.000122                                                                                                     & \multicolumn{1}{r|}{0.023}                                                                         & 0.026                                                                                         & 0.01                                                                                        & \multicolumn{1}{r|}{0.01}                                                 & \multicolumn{1}{r|}{2.3}                                                         & 2.5                                                                              \\ \hline
movielens-t-i                                & 7,601                                     & 16.528                                    & 71,154                                    & 1.1E-03                                                                                      & 140,266                                                                                                 & 0.53                                                                                                         & \multicolumn{1}{r|}{5.34}                                                                          & 3.07                                                                                          & 0.47                                                                                        & \multicolumn{1}{r|}{1.13}                                                 & \multicolumn{1}{r|}{11.36}                                                       & 6.53                                                                             \\ \hline
movielens-u-i                                & 7,601                                     & 4,009                                     & 55,484                                    & 3.6E-03                                                                                      & 2,365,457                                                                                               & 1.27                                                                                                         & \multicolumn{1}{r|}{10.97}                                                                         & 6.67                                                                                          & 0.85                                                                                        & \multicolumn{1}{r|}{1.49}                                                 & \multicolumn{1}{r|}{12.91}                                                       & 7.84                                                                             \\ \hline
movielens-u-t                                & 16,528                                    & 4,009                                     & 43,760                                    & 1.3E-03                                                                                      & 166,380                                                                                                 & 0.30                                                                                                         & \multicolumn{1}{r|}{0.95}                                                                          & 0.635                                                                                         & 0.1                                                                                         & \multicolumn{1}{r|}{3}                                                    & \multicolumn{1}{r|}{9.5}                                                         & 6.36                                                                             \\ \hline
UCforum                                      & 522                                       & 899                                       & 7,089                                     & 3.0E-02                                                                                      & 16,261                                                                                                  & Fault                                                                                                        & \multicolumn{1}{r|}{0.09}                                                                          & 0.09                                                                                          & 0.02                                                                                        & \multicolumn{1}{r|}{X}                                                    & \multicolumn{1}{r|}{4.5}                                                         & 4.66                                                                             \\ \hline
Unicode                                      & 614                                       & 254                                       & 1,255                                     & 1.6E-02                                                                                      & 460                                                                                                     & 0.00094                                                                                                      & \multicolumn{1}{r|}{0.033}                                                                         & 0.052                                                                                         & 0.004                                                                                       & \multicolumn{1}{r|}{0.24}                                                 & \multicolumn{1}{r|}{8.25}                                                        & 13.02                                                                            \\ \hline
\end{tabular}%
}
\end{table*}

\subsection{Fine-grained Parallelism in Independent Subtrees}
\label{sec:fine-grained}

Besides the coarse-grained parallelism,
there are numerous opportunities to parallelize the inside a coarse-grained task (subtree).
In this section, we design three fine-grained optimizations on GPUs: \emph{candidate selection with early stops}, \emph{reverse scanning on maximality checking and maximal expansion} and \emph{lookup table for fast querying}.

\textbf{Candidate selection with two early stops.} As we have discussed in Section \ref{sec:techniques}, 
the selecting order of the candidate can significantly avoid unsuccessful searching and reduce imbalance among recursion subtrees.
Prior works \cite{zhang2014finding, das2019shared} calculate the vertices' priority and sort them in $P$ one time before the recursive procedure.
This method works in most CPU cases but is infeasible in our case where the $P$ set uses the compact array representation mentioned in the previous section.
The order of $P$ indicates which recursive level the procedure is processing at;
hence, arbitrarily sorting $P$ at each level leads to error behavior.
The only way to maintain the same degeneracy order is to search the vertex with the highest priority in the candidate set $P$ at every recursion.
However, the vanilla searching procedure which has to go through the $P$ is time-consuming.
To solve this inefficiency, 
we design the parallel searching optimization with \emph{two early stop points} to avoid the redundant checking on $P$ set, shown in Figure \ref{fig:candidate_selection}. 
Our approach is suitable for any ordering methods mentioned in Section \ref{sec:techniques}.
For the sake of clarity in explanation, 
we take the degeneracy order used in iMBEA \cite{zhang2014finding} for example.
It selects the vertex $v$ which makes the $L'$ shrink the most; 
namely, the smallest $|N(v) \cap L|$.
In this case, we maintain three values during the process: \texttt{cnt}, \texttt{last\_min}, and \texttt{cur\_min}.
The \texttt{cnt} counts the $|N(v) \cap L|$;
the \texttt{last\_min} is the last minimum value of the vertex selected as the last candidate,
while the \texttt{cur\_min} is the temporally minimum value found in the current level.
We assign the whole thread block to compute vertex by vertex.
Each thread inside a thread block is responsible for a neighbor $u$. 
Then, we utilize the look-up table $LT^{u}_{L}$ for determining if $u$ is in $L$ and atomically increments \texttt{cnt}.
Whenever, \texttt{cnt} exceeds \texttt{cur\_min}, 
we can bypass the computation of $v$ since there is another vertex whose value is smaller than $v$.
It is the first early stop point.
After iterating all the neighbors of $v$, 
if \texttt{cnt} equals to \texttt{last\_min}, 
we can denote this vertex as the highest priority of this level and stop iterating other vertices.
It is the second early stop point.
The early-stop approach proves effective on real-world datasets because the degree of the vertices within these datasets typically falls within a relatively small range.

\textbf{Reverse scanning on maximality checking and maximal expansion.} These two steps have similar operations but operating on different sets ,$Q$ and $P$. 
Fundamentally, they iterate all vertices from a set and check if the neighbors of each vertex connect to the $L'$ set.
If a neighbor vertex fully connects to $L'$, we will detect the violation in maximality checking, or add the neighbor vertex to $R'$ in maximal expansion. 
In our observation, the size of $L'$, which significantly shrinks following the processing of the $1^{st}$ level subtree,
is much smaller than the size of the $Q$ and $P$ sets when we descends the searching tree.
Besides, the shrinking phenomenon becomes even more pronounced when we apply the order of candidate selection discussed in Section \ref{sec:techniques}.
Hence, we achieve the maximality checking and maximal expansion reversely;
namely, We iterate the vertices in the $L'$ set rather than the vertices in the $P$ and $Q$.
For the clear explanation, we use the maximality expansion for example.
Inspired by the parallel intersection computation of ParMBE \cite{das2019shared}, 
which utilizes a hash map to record the number of common neighbors with the vertex as the key,
we maintain a buffer consisting of the vertex $v$ from $P$ set and the value $y$ representing the number of common neighbor $|N(v) \cap L'|$.
With this buffer, we can move the vertex in $P$ to $R'$ if its $y$ equals to $|L'|$, 
while the vertex with $y > |L'|$ belongs to $P'$ (line: 15-18 in Algorithm \ref{algo:MBEA}).
In addition, we adapt the two-level parallelism on the scanning step. 
Specifically, we assign a warp for a vertex $v$ in $L'$, 
while threads in a warp parallely visit the neighbors of the vertex;
namely, a thread checks a neighbor vertex if it is in $P$ set; 
then, the thread will atomic add or increment the value on buffer with the index $v$.

\textbf{Lookup table for fast querying.} In the MBEA, there are numerous operations which we have to query if a vertex is in the given set. 
For instance, in candidate selection, we have to check if the neighbor vertices of the vertex $v$ is in $L'$ set to calculate the size of $|N(v) \cap L'|$.
In addition, the maximality checking and maximal expansion require the same query. 
However, we observed that performance this query on our compact array is really time-consuming on GPUs.
Traditionally, we have to linearly iterate all the vertices in a set, which takes $O(d)$ where $d$ is the size of set. 
Even we apply the parallel searching on the process, the complexity is still $O(log_{t}d)$ where $t$ is the number of threads we assign. 
To address this inefficiency, we design the \emph{lookup table} upon the compact array shown in Figure \ref{fig:compact_stack}, \ref{fig:candidate_selection}.
Fundamentally, each compact array equips up a lookup table whose size equals to the corresponding compact array.
The index of the table represents the vertex ID of the vertices in the array, 
while the value of the table is the position of the vertex in the compact array.
For instance, $LT^{u}_{P}[0] = 8$ is because $P[8]$ stores the vertex $0$ in Figure \ref{fig:compact_stack}.
The lookup table updates following by the change of the corresponding compact array.
With this table, we can reduce the complexity of the vertex query operation to $O(1)$ by comparing the value stored on the table and the value of the level pointer.

\subsection{Theoretical Analysis of cuMBE}
\label{sec:theoretical_analysis}

\textbf{Space complexity analysis.} The compact array of cuMBE reduces the memory requirement of the recursive search of MBE from $O(|V+U| \times T \times H)$ to $O(|V+U| \times 2 \times T )$ by replacing the time-consuming dynamic memory allocation of a new set to a disjoint array sized $O(|V+U|)$ with four pointers per searching level to represent the changes in the four sets, $L, P, Q$, and $R$, 
where $T$ is the number of workers, and $H$ represents the maximum height of the searching forest shown in Figure \ref{fig:recursion_tree}. 

\textbf{Time complexity analysis.} The execution time model of a MBE can be represented by the following equation:

\begin{equation}
    time = max_{T} (\sum_{i=0}^{W_t}(A_i + B_i + C_i + E_i))
\end{equation}

\noindent 
In this equation, $T$ represents the number of workers (thread blocks), and $W_t$ represents the permutation number (or workload) handled by worker $t$. $A, B, C$, and $E$ represent the execution time of candidate selection, $L'$ construction, maximal checking, and maximal expansion in the i-th permutation, respectively.
There are two main directions to improve the overall execution time: \emph{distributing the workload evenly and enhancing the efficiency of each worker in executing a permutation}. 
Our work-stealing mechanism aims to achieve the former, while the three fine-grained optimizations are designed for the latter.

The early stop mechanism on candidate selection avoids scanning the entire $P$ set to calculate the neighbor size of every candidate vertex. 
Instead, we design two breakpoints, which are clearly described in Section 
\ref{sec:fine-grained}, which changes the time complexity of $A$ from $\Theta(|P|*D(P))$ to $\Omega(1)$ and $O(|P|D(P))$, where the $D(P)$ represents the maximum degree of vertices in set $P$.
In addition, the lookup table optimization accelerates the query procedure from $O(S)$ to $O(1)$, where S is the size of the given set.
Lastly, the work-stealing focuses on amortizing total workload $W$ equally among all workers, as the worker with the most workload will determine the overall execution time of a GPU. We choose the work-stealing approach because of its greedy and dynamic characteristics. It requires minimal extra hyperparameters and has been proven to have good improvements.

\section{Evaluation}
\label{sec:evaluation}

In this section, we first introduce our experiment setup. 
Then, we evaluate the performance between our design and the state-of-the-art approaches. 
We will discuss the overall speedup compared to them,
then we point out that the improvement results from the workload balance caused by our k-level work-stealing algorithm and the three fine-grained parallelism optimizations. 
Lastly, we analyze our design and show the its limitations and future works.

\begin{figure*}
    \centering
    \includegraphics[width=0.97\linewidth]{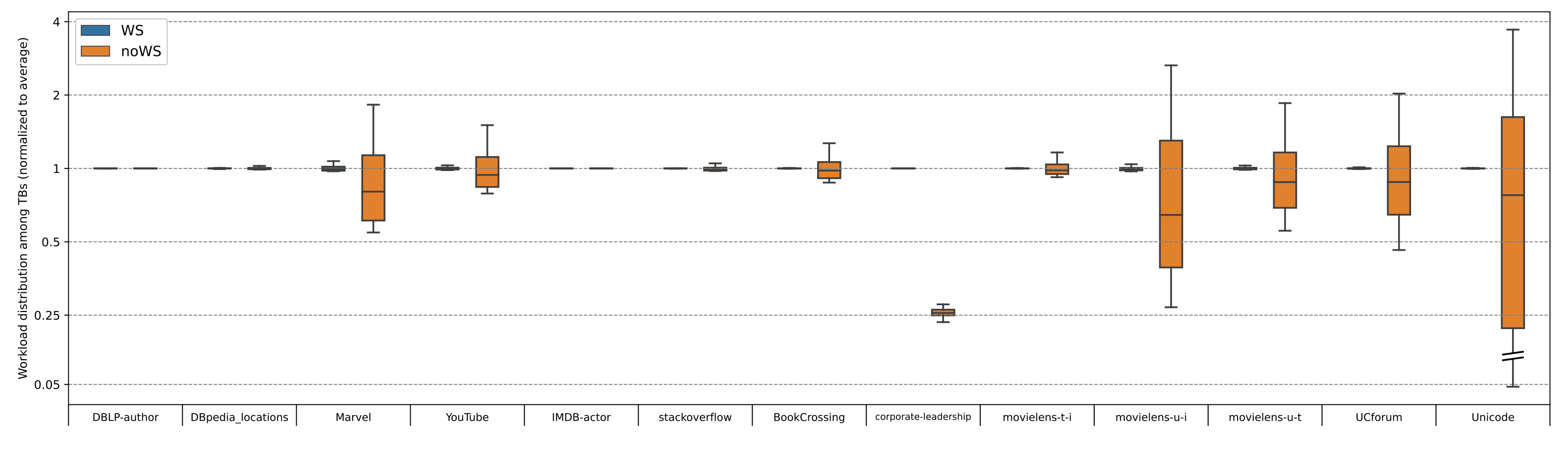}
    \caption{The load distribution among thread blocks across different datasets on a NVIDIA RTX 3090 GPU.}
    \label{fig:workload}
\end{figure*}

\subsection{Experiment Setup}

\noindent
\textbf{Implemented algorithms.} To the best of our knowledge, our work, \emph{cuMBE}, is the first MBE algorithm on GPUs;
hence, there is no GPU baseline.
We select two state-of-the-art algorithms, a serial and a parallel algorithm on CPUs for comparison.

\textit{\underline{ooMBE \cite{ooMBE}.}} This is the state-of-the-art serial MBE algorithm with multiple pivot pruning and the novel candidate selection policy. We use and fix the compiling bug of the code provided by the author of ooMBE \cite{ooMBE-code}.

\textit{\underline{ParMBE \cite{das2019shared}.}} This is the parallel MBE algorithm designing for CPUs. It applies the Intel's Thread Building Block (TBB) \cite{pheatt2008intel} on the operation of iMBE \cite{zhang2014finding} with rank pruning optimization. 
Since the project is closed-source, we faithfully re-implement ParMBE .

\textit{\underline{cuMBE.}} We use the function \texttt{grid.sync()} provided by NVIDIA to synchronize thread blocks and fuse the whole MBE procedure to a single kernel. 
We configure a TB as 512 threads and there are total 246 TBs in our GPU platform (RTX 3090)

\noindent
\textbf{Datasets.} We use the same datasets used by ooMBE and ParMBE which are shown in Table \ref{table:performance}.

\noindent
\textbf{Measuring machines.} We evaluate the execution of the above algorithms on three machines. 
The first machine equips with an NVIDIA RTX 3090 GPU attached to Intel i9-10900k CPU @ 3.7GHz for GPU evaluation. 
The second and third machines are the virtual machines rented from Amazon EC2 for evaluating CPU performance.
We choose a dual-CPU processor with Intel Xeon Platinum 8375C (Ice Lake) CPU @ 2.9 GHz, which has up to 64 cores (128 threads) and 256 GB memory.
Another machine is a dual-CPU processor with AMD EPYC 7R13 (Zen 3) CPU @ 2.7 GHz, which has 96 cores (192 threads) and 256 GB memory for completeness. 
These two machines are the most two advanced instances provided by Amazon EC2 service.
The serial ooMBE runs on the Intel platform. 

\subsection{Performance}
\label{sec:performance}

Table \ref{table:performance} presents the performance evaluation results of the MBE algorithms. 
In this table,  we have chosen large datasets represented by the first seven columns, from \emph{DBLP-author} to \emph{BookCrossing}, and smaller datasets depicted by the subsequent six columns, from \emph{corporate-leadership} to \emph{Unicode}. We calculated the edge density using the formula $\frac{2*|E|}{|L| * |R|}$ to indicate dataset sparsity.
Our analysis shows that cuMBE surpasses ParMBE in most datasets, except for \emph{DBLP-author}. On the Intel platform, cuMBE averages a speedup of 7.1x and 5.5x on the AMD platform. 
However, cuMBE consumes a significant amount of time retrieving coarse-grained tasks in \emph{DBLP-author}. Section \ref{sec:ablation} delves deeper into the reasons behind cuMBE's suboptimal performance on \emph{DBLP-author}.
Compared to ooMBE, cuMBE achieves an average speedup of 6.1x across all datasets. The specific speedup varies with the dataset. ooMBE's unique candidate selection ordering shows a preference for sparse bipartite graphs, 
leading to strong performance on datasets such as \emph{DBLP-author}, \emph{DBpedia\_locations}, \emph{IMDB-actor}, and \emph{stackoverflow}, 
all with edge densities below $10^{-4}$. 
cuMBE's reduced speedup on these datasets stems from its inability to leverage all available parallelism.
Lastly, datasets with a high $\frac{nMB}{|E|}$ ratio, 
where $nMB$ denotes the number of maximal bicliques, 
see significant performance boosts with cuMBE. 
Examples include \emph{Marvel}, \emph{YouTube}, \emph{BookCrossing}, and \emph{movielens-u-i}.

\subsection{Workload Analysis}
\label{sec:workload_anaylsis}

Figure \ref{fig:workload} illustrates the workload distribution across 13 datasets. 
The X axis represents different datasets, 
while the Y axis is the execution time normalized to the average execution time of each dataset.
We recorded the execution of all the 246 thread blocks on RTX 3090 for each dataset. 
The blue and orange bars are the cuMBE implementation with and without 2-level work-stealing mechanism, respectively. 
The bar shows the minimum, maximum, lower quartile, upper quartile and median execution among thread blocks.  
At a high level, our work-stealing mechanism can significantly alleviate the imbalance across every datasets. 
The standard deviation range on each dataset shrinks from $5.2 \times 10^{-4} \sim 2.9$ to $3.3 \times 10 ^{-4} \sim 3.1 \times 10^{-2}$.
The small datasets benefit the most from our work-stealing mechanism. 
However, in some extremely tiny datasets, for instance \emph{corporate-leadership}, \emph{UCforum} and \emph{Unicode}, the overhead of 2-level work-stealing is much larger than the improvement of it, 
which can be clearly observed in Figure \ref{fig:breakdown}.
In addition, we point out that the 2-level work-stealing is enough for all the datasets since each TB seems to have equal workload.
The overhead of more fine-grained work-stealing mechanism might outweigh the advantages of being more balanced. 
It can be supported by Figure \ref{fig:breakdown} that the portion of the idling time becomes really small after we used the 2-level work-stealing mechanism.

\subsection{Ablation Study}
\label{sec:ablation}

\begin{figure*}[t]
    \centering
    \includegraphics[width=0.97\linewidth]{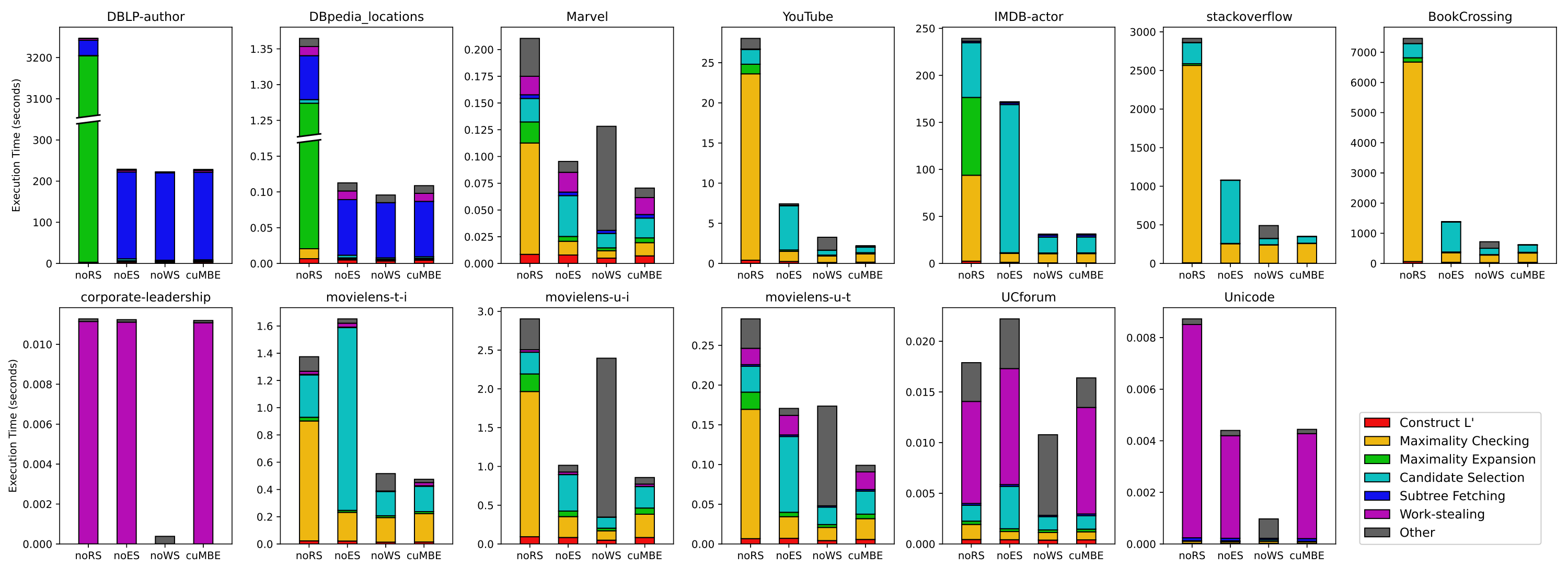}
    \caption{The execution time breakdown of cuMBE with different combinations of fine-grained parallelism optimizations across various datasets. Each bar represents the execution time with one optimization removed: \emph{noRS}: remove the reverse scanning on maximality checking and expanding; \emph{noES}: remove the early stop technique on candidate selection; \emph{noWS}: remove the work-stealing technique on second-level subtree. \emph{cuMBE}: applying all the three optimizations.}
    \label{fig:breakdown}
\end{figure*}

In this section, we evaluate the three optimizations related to fine-grained parallelism, as discussed in Section \ref{sec:fine-grained}.
Since we are pioneering the application of MBE on GPUs with cuMBE, 
we delve into the effectiveness of our specific design.
Figure \ref{fig:breakdown} provides a breakdown of execution time across various datasets,
captured using NVIDIA CUDA's \texttt{clock()} function.
Each bar in the figure represents a unique combination of optimizations.

We segment the entire MBE algorithm's execution time into eight components:
\emph{candidate selection}, \emph{L' construction}, \emph{maximality checking}, \emph{maximality expansion}, \emph{subtree fetching}, \emph{work-stealing} and \emph{others}, 
The first four components align with the MBE algorithm described in the background section, 
Subtree fetching and work-stealing relate to time spent on coarse-grained parallelism and the work-stealing algorithm, respectively. 
Specifically, the subtree fetching part consists of the two overhead:
atomic operations on contending coarse-grained tasks against other TBs, 
and the memory latency of copying from $P_{g}$ to local $P_{p}$.
The work-stealing segment comprises the sorting time of $P_{g}$ on victim thread blocks and the time of selecting a victim thread block.
The last part, Others, includes the kernel launching time and the idling time of each thread block.
The category labeled others encompasses kernel launching and thread block idling times.

% Describe how different optimizations affect the execution time
At a glance, the three fine-grained parallelism optimizations--reverse scanning, early-stop, and work-stealing mechanism--do indeed enhance execution time for most datasets.
However, work-stealing tends to increase overhead for smaller datasets, 
diminishing its utiliy in those cases.
On larger datasets, the idling time comprises about 16.8\% of the total execution time without the work-stealing,
highlighting the sufficiency of 2-level work-stealing for balancing the workload.
The work-stealing mechanism can achieve average 2.18x and 1.64x speedup on large datasets and small dataset, respectively.
It even gains 1.7x and 2.7x speedup on Marvel and movielens-u-t, whose workload is much more imbalanced.
Although enhancement of the balancing is limited,
our work-stealing mechanism avoids the worst-case imbalanced situation on any large datasets.

In contrast, 
reverse scanning on maximality checking and expansion brings remarkable enhancement.
It gains 2x to 11.3x speedup on most of the datasets, 
except for the small ones.
The early-stop mechanism on candidate selection particularly benefits datasets with extensive candidate sets, such as IMBD-actor and movielens-t-i.
Besides these broad perspectives, we make several observations and will discuss in the following.

% DBLP and DBpedia's poor perfomrance
The first observation is that cuMBE spends extensivve time on fetching the subtree in DBLP-author and DBpedia\_locations.
The subtree fetching, which is no need in the prior CPU-based serial or parallel works, accounts for more than 80\% of the total time on both DBLP-author and DBpedia\_locations in cuMBE configuration.
It is the reason why cuMBE is overwhelmed by ooMBE and ParMBE on those datasets, as we discussed the weakness in Section \ref{sec:performance}.
This phenomenon correlates with the datasets' community-rich structures.
We can easily observe that the time spending on maximality expansion is much longer than the time on maximality checking in noRS configuration on DBLP-author and DBpedia\_locations, 
which indicates that the biclique expands to maximal in the early level and returns from the recursion right after the expansion; 
hence, there is few need to check the maximality.

Based on this observation, we can infer that there are numerous communities on these datasets.
Specifically, when a maximal biclique is expanded and found, few vertices can be further added to the biclique to form another maximal biclique.
This hypothesis can be supported by the characteristic of these datasets. 
For instance, DBLP-author is a authorship network, where the $L$ set consists of authors and the $R$ set comprises publications.
Since the network represents the connection between authors and publications, 
it is really sparse and comprises numerous communities.
Because of the clustering, the height of subtrees is generally small,
which means a TB spends relatively more time on copying subtree than others.
Although the reverse scanning and the lookup table optimizations have  significantly reduced the time on maximality expansion and checking, there remains an opportunity to improve memory transfer latency during subtree fetching.

% Observation 6: the correlation between optimizations
Additionally, we noticed several inter-dependencies between optimizations.
First, execution time for maximality checking and expansion directly impacts subtree fetching time. 
In DBLP and DBpedia\_locations, which spend remarkable time on subtree fetching, 
the \emph{noRS} configuration has less time spending on subtree fetching compared to other configurations.
Since \emph{noRS} spends much more time on maximality expansion, 
the contention on the global $P$ set is reduced.
Moreover, the latency of copying global $P$ to local $P$ on each TB can be better hidden with the warp scheduling once the execution time is longer.
The built-in warp scheduler in a GPU context-switch to another warp when a warp is waiting for memory transferring. 
These two reasons that make the portion of subtree fetching shorter.

Our reverse scanning optimization also synergizes well with early stopping, resulting in more efficient candidate selection.
The candidate selection time for \emph{noRS} is, on average, 2.8 times longer than that of \emph{noWS} and \emph{cuMBE} across the datasets--YouTube, IMDB-actor, stackoverflow, and BookCrossing.
This is because our reverse scanning on maximality expansion will move the vertex, which has at least one common neighbor with $L'$, to the beginning of the next-level candidate set $P'$ (In Algorithm \ref{algo:MBEA} line 17 - 18);
hence, this perturbed $P'$ set has better performance under the early-stop optimization since it can get the candidate more early than non-perturbed one. 

In summary, although work-stealing has its limitations, both reverse scanning and early stopping have emerged as powerful strategies for performance enhancement in cuMBE. Our observations underscore the need for continued investigation into reducing memory latency and further optimizing the balance between various algorithmic components.

\section{Related Works}

MBE has been widely studied for decades.
Alex et.al \cite{alexe2004consensus} used consensus algorithms to find all maximal bicliques. 
Liu et.al \cite{liu2006efficient} is the first depth-first search-based approach on MBE, called MineLMBC.
Zang et.al's iMBE algorithm \cite{zhang2014finding}, which improves MineLMBC by further reducing the next-level candidate set.
Because iMBE is the most common used algorithm to develop various optimizations on it, for instance PMBE \cite{abidi2020pivot} and ooMBE \cite{chen2022efficient},
we referenced iMBE to design cuMBE.

There are also numerous prior works related to MBE. 
Yang et.al \cite{yang2023p} introduced BCList++ to effectively find the (p, q)-biclique, where the the p and q valuables are given by users. 
Ye et.al \cite{ye2023efficient} targeted on the (p, q)-biclique counting, which reduces the exponential blowup in the enumeration space by using edge-pivoting technique, called EPivoter.
Maximum balanced biclique, whose $L$ and $R$ set size are equal, is also an important research area \cite{chen2021efficient, zhao2022finding}. 
All of the above algorithms are serial.
Our cuMBE is the first biclique-related algorithm utilizing GPUs for accelerating the searching progress.

Maximal clique enumeration (MCE) is closed research topic to MBE.
The clique is the complete subgraph in general graph. 
Despite numerous prior works have proposed the parallel version of MCE on CPUs \cite{schmidt2009scalable, blanuvsa2020manycore, dasari2014pbitmce, das2020shared, yu2019memory} or GPUs \cite{wei2021accelerating, almasri2022parallelizing, alusaifeer2013gpu, jayaraj2016gpu}, 
the reduction from MCE to MBE is infeasible.
G{\'e}ly et.al \cite{gely2009enumeration} attempted to use MCE on the extended bipartite graph, 
but the additional edge number increases dramatically, leading to the performance degradation;
hence, designing the MBE algorithm on GPUs is necessary and essential.
To the best of our knowledge, cuMBE is the first MBE on GPUs.

Our optimizations of cuMBE on GPUs mentioned in Section \ref{sec:fine-grained} are inspired by the prior works of graph data mining.
The two-level parallelism technique has been widely adapted on graph pattern mining \cite{chen2022efficient, chen2021sandslash, kim2020batch} and graph neural network \cite{fu2022tlpgnn}.
We use inter-warp parallelism for iterating the vertex in a set, while the intra-warp parallelism is used to achieve data parallelism on visiting the neighbors of the given vertex.
There are also numerous works trying to alleviate the workload imbalance on GPUs \cite{almasri2022parallelizing, merrill2012scalable, davidson2014work, fu2022tlpgnn, hsieh2023decentralized}.
The general method to reduce the inter-warp imbalance is to reassign tasks on warps \cite{almasri2022parallelizing, fu2022tlpgnn, hsieh2023decentralized},
while the intra-warp imbalance can be reduced by re-configuring the warp size \cite{merrill2012scalable, davidson2014work}.
Our k-level work-stealing optimization, which addresses the inter-block workload imbalance is similar to the former approach, 
yet we dynamically enable the task to be more fine-grained, 
which can achieve more balancing effect.
The intra-warp imbalance still remains enormous researching space.

\section{Conclusion}

We present cuMBE, to the best of our knowledge, the first parallel MBE algorithm on GPUs that eliminates the dynamic memory allocation overhead and memory explosion caused by the traditional recursion-based MEB algorithm.
We utilize hybrid parallelism to accelerate MBE on GPUs.
We assign a thread block for exploring the searching space in depth-first manner.
During the searching progress, 
we design three intra-block optimizations, early-stop mechanism, reverse scanning and lookup table, for further improving the performance.
In addition, a k-level work-stealing mechanism reduces the workload imbalance among thread blocks.

Our evaluation shows that cuMBE significantly outperforms the state-of-the-art serial and parallel MBE algorithm on CPUs. 
The workload analysis also supports our work-stealing mechanism can successfully alleviate workload imbalance. 
Lastly, we analyze cuMBE and identify its weaknesses and limitations with a detailed ablation study.

\bibliographystyle{IEEEtran}
\bibliography{refs}

\end{document}